\documentclass{iopart}
\usepackage{graphicx,latexsym,iopams,hyperref,amsfonts}

\begin{document}

\title{Focus on topological quantum computation}
\author{Jiannis K. Pachos}
\address{School of Physics and Astronomy, University of Leeds, Leeds, LS2 9JT, United~Kingdom} 
\ead{j.k.pachos@leeds.ac.uk}
\author{Steven H. Simon}
\address{Rudolf Peierls Centre for Theoretical Physics, University of Oxford, OX1 3NP, United Kingdom} 
\ead{s.simon1@physics.ox.ac.uk}
\date{\today}

\begin{abstract}
Topological quantum computation started as a niche area of research aimed at employing particles with exotic statistics, called anyons, for performing quantum computation. Soon it evolved to include a wide variety of disciplines. Advances in the understanding of anyon properties inspired new quantum algorithms and helped in the characterisation of topological phases of matter and their experimental realisation. The conceptual appeal of topological systems as well as their promise for building fault-tolerant quantum technologies fuelled the fascination in this field. This `focus on' brings together several of the latest developments in the field and facilitates the synergy between different approaches.

\end{abstract}

The great promise of quantum computers has to be balanced against the great difficulty of actually building them. Foremost among the difficulties is the fundamental challenge of defeating decoherence and errors. Small improvements to current strategies may not be sufficient to overcome this problem; radically new ideas may be required. Topological quantum computation is precisely such a new and different approach \cite{Kitaev97}. It employs many-body physical systems with the unique property of encoding and processing quantum information in a naturally fault-tolerant way.

Research on topological quantum computation has become a highly interdisciplinary field, with frontiers in physics, mathematics, and computer science. Moreover, advances in the theoretical understanding of abstract topology \cite{Jones85}, in physical realisations of topological matter \cite{Nayak}, and in computational paradigms \cite{Freedman03} have been closely interrelated, with developments in one area strongly influencing the others. Recent years have witnessed significant theoretical and experimental developments. These include major experimental and theoretical advances in fractional quantum Hall systems that support the existence of non-Abelian anyons \cite{Moore-Read} -- the building block of topological quantum computation -- as well as the predication and experimental discovery of novel spin-orbit systems such as topological insulators \cite{Hasan12}. 

The aim of this focus issue is to bring together the latest developments in a variety of areas with the goal of promoting new results to a wider community of scientists and advancing the synergy between different approaches. To a large extent we believe this has been a success. 

One of the dominant directions of the field has been towards better understanding topological matter by investigating phase transitions between topological phases --- how one phase condenses out of another, often in some sort of confinement transition.  Loop models, as the simplest of all topological phases, were studied by Monte-Carlo methods \cite{Herdman}.   Semi-analytic high order perturbation theory methods were applied to study transitions in ${\mathbb{Z}}_N$ topological phases \cite{Schulz}.   In some cases powerful purely analytic approaches could be applied to study phase transitions in detail \cite{Burnell,Bombin,Ludwig}.    There was substantial focus on identifying order parameters for such topological phase transitions to try to draw analogies with the conventional Landau theory of phase transitions \cite{Bais,Gregor}.   Despite this progress, and although some general principles are now established \cite{BaisSlingerland}, a unified theory of the details of phase transitions between topological 
phases is still absent and remains an active topic of research. 

Another emerging direction in the field has been the study of Majorana fermions and the physical systems that may harbour them.  In the past few years, several proposals were made for creating analogs of chiral p-wave superconductors using spin-orbit coupled superconductors, which would host Majorana zero modes at defects such as vortices. Braiding these defects would result in non-Abelian operations on a degenerate ground state \cite{Read} (see \cite{AliceaReview} for a review of these ideas). Indeed, this development became such an active field that it called for a separate focus issue \cite{Majorana}.  However, before that issue was created many groundbreaking works on the topic ended up in the current focus issue.   One direction is in determining the parameters necessary for obtaining topological superconductivity in spin orbit coupled systems \cite{Tewari}.  Another direction is the proposal of experiments (in this case Ettingshausen effect experiments \cite{Hou} or quantum point contact tunnelling experiments \cite{Wimmer}) that might establish the existence of Majorana modes in these systems. 
 Yet another direction is in the development of device designs to read out \cite{Hassler1,Hassler2}  or manipulate \cite{vanHeck} such Majoranas qubits.     Majoranas, as perhaps the simplest non-Abelian object, were also studied in several other contexts including phase transitions \cite{Burnell}, fractional quantum Hall effect \cite{Clarke}, the Kitaev Honeycomb model \cite{Kitaev05,Ville,Ahmet,Kells}, Hanbury-Brown-Twiss experiments, \cite{Bose} and quench dynamics of spin chains \cite{Sen}.  It is certain that the study of Majorana physics will be a main direction of the field in the future. 

A perennially dominant direction since the inception of the field has been the study of fractional quantum Hall effects \cite{Stormer,Nayak}.  This is perhaps not surprising being that quantum Hall effect in semiconductors remains the only physical system convincingly established as nontrivial topological matter.  In this issue quantum Hall physics was also discussed in other physical realisations ranging from mono-layer and bilayer graphene \cite{Abanin} to cold atoms in non-Abelian gauge fields \cite{Palmer,Estienne}.  The presence of extra degrees of freedom, such as spin and valley,  over the conventionally studied spinless electrons, gives the freedom to have richer physics, including  charged spin textures \cite{Romers}.   In another direction, a nice theoretical advance (the extension of a many year project by some of the same authors and others) was the development of a conformal field theory approach to the quantum Hall hierarchies \cite{Soursa}.  As in the broader field of condensed matter, 
understanding the patterns of entanglement of fractional Hall states has becoming an increasingly important and interesting topic \cite{Sterdyniak,Friedman}.   

While several experimental works have recently been performed attempting to demonstrate non-Abelian quasiparticles in the fractional quantum Hall effect \cite{Willett,Kang}, the interpretation of these experiments is very controversial, and this remains a forefront of research.   Several works in the current issue were aimed at either providing other methods to make this demonstration, for example by measuring tunnelling spectra \cite{Hu}, or were aimed to examine additional physics of these experiments (in this case the Zeno effect \cite{Clarke}, or disorder in quasiparticle lattice \cite{Kraus}) not previously considered in predictions.  In our entire focus issue, disappointingly, only a single submitted manuscript was actually a real experiment \cite{Choi}, and this, although a nice experiment demonstrating Aharonov-Bohm oscillations in quantum Hall samples, remains quite a distance from the desired result that would convincingly show non-Abelian physics.

Back to the theoretical front we note that perhaps one of the most studied models has been the Kitaev honeycomb model \cite{Kitaev05}.   Unsurprisingly this continued to provide fertile ground for further thought.  Within this framework the interactions between non-Abelian vortices and the nucleation of new topological phases was studied \cite{Ville} and the braiding statistics were evaluated explicitly to very high accuracy \cite{Ahmet}.  A square-octagon variant of the Kitaev honeycomb model was shown to exhibit a rather remarkably rich phase diagram \cite{Kells}. 

Another of the most celebrated models of topological matter is the toric code model \cite{Kitaev97}. In one work, the thermal stability of this model was examined \cite{Abbas}.  Another work considers how this model might actually be simulated using trapped ions \cite{Muller}.  Since its inception, the toric code model has since been generalised to ${\mathbb Z}_N$ models \cite{Schulz}, non-Abelian discrete gauge theories \cite{Bombin}, and finally to their most general form, the Levin-Wen models \cite{LevinWen,Burnell,Burnell2,Ludwig,Ardonne}.   One work in this issue geometrically reinterpreted the Levin-Wen partition functions as knot invariants \cite{Burnell2}.  Another work explores (at least in one dimension) generalising to theories based on non-unitary conformal field theories \cite{Ardonne}.

On the more computational side of research, one of the key ideas that launched the field was the realisation that evaluation of the Jones polynomial, a ``hard'' problem for a classical computer, might not be ``hard'' for an anyon computer \cite{Freedman98}.   Indeed, evaluation of Jones polynomials \cite{Freedman03,Aharonov} is perhaps  one of the most natural algorithms for a topological quantum computer. In this issue, extensions of this line of thought have resulted in advances in determining the precise complexity of this quantum algorithm \cite{AharanovArad} and in its relationship to Khovanov homology --- a natural ``categorification'' of the Jones polynomial  \cite{Kauffman}.

The quantum gates produced by anyonic braiding are very specific, tightly connected to the statistical properties  of the anyons. In this focus issue, it was shown that no computationally universal anyon system can be built where two qubit gates do not have finite leakage errors \cite{Ainsworth}.  This further draws attention towards the necessity of approximating quantum gates to high accuracy (even though they must remain imperfect).  One approach pursued towards designing such high accuracy gates is to use hashing techniques \cite{Burrello}.  Finally, ideas from cluster state quantum computation have been applied to the realm of topological models \cite{Horsman,Brown}.

As witnessed from these contributions it is clear that the field of topological quantum computing, and topological matter in general, has continued to progress at a remarkable rate.   Despite these advances, several fundamental questions still remain open.  Arguably, two main directions can be identified: building systems where non-Abelian statistics can be conclusively measured and finding a topologically inspired quantum memory that can successfully combat physically induced errors. It is the hope and vision of the editors that bringing together methods and techniques from wide range of fields can eventually enable the construction of topological quantum computation.

\end{document}